\documentstyle[preprint,aps,prabib]{revtex}
\begin{document}
\title{Domain with Noncompactified Extra Dimensions in
Multidimensional Universe with Compactified Extra Dimensions}
\author{Dzhunushaliev V.D.
\thanks{E-mail address: dzhun@freenet.bishkek.su}}
\address {Freie Universit\"at Berlin, Fachbereich Physik, 
Arnimalleee 14, 14195 Berlin, FRG;\\ 
Dept. of Theor. Phys. Kyrgyz State National University\\  
Bishkek, 720024, 
Kyrgyzstan}

\maketitle

\begin{abstract}
It is supposed that in our Universe with compactified 
extra dimensions (ED) the domains exist with noncompactified 
ED. Such domain can be a wormhole-like solution in multidimensional 
gravity (MD), located between two null surfaces. With the 
availability of compactification mechanism this MD domain 
can be joined on null surfaces with two black holes filled 
by gauge field. Solution of this kind in MD gravity 
on the principal bundle with structural group SU(3) 
is obtained. This solution is wormhole-like object located 
between two null 
surfaces $ds^2=0$. In some sense these solutions are dual to 
black holes: they are statical spherically symmetric 
solutions under null surfaces whereas black holes 
are statical spherically symmetric solutions outside 
of event horizon.
\end{abstract}

\par
Name:	   	Dzhunushaliev Vladimir
\par
Postal address:	mcr.Asanbai, d.25, kv.24, 720060, Bishkek, 
Kyrgyzstan
\par
Telephone:	007 (3312) 46-57-45
\par
E-mail:		dzhun@freenet.bishkek.su
	
\newpage

\section{Introduction}

Currently have been understood that 
an important feature of modern Grand Unified Theories 
(GUT) is its multidimensionality (MD). In these theories our Universe 
is MD Universe with compactified extra dimensions. In 80-th 
years the MD gravity on fibre bundles with fibres=symmetrical 
space had been investigated intensively (see, for example, 
\cite{wit}-\cite{coq}). In these works had been established 
that above mentioned MD gravity is equivalent to 4D gravity + 
gauge field + scalar field. The multidimensionality is 
a very serious component for almost all recent GUT. 
In these theories the superfluous extra dimensions (ED) are 
frozen and contracted to very small sizes. We can suppose, nevertheless, 
that there are some regions with noncompactified ED in 
our MD Universe. This can be true near to singularity which 
make up the pointlike elementary particle and cosmology singularity. 
In first case such domains with noncompactified ED can be found 
under event horizon.
\par
Now the cosmological solutions, describing the properties of 
P-brains, interacting with MD gravity are investigated 
intensively \cite{bron}. Also MD cosmological models 
with MD spacetime $M=M_0\times\prod_{i=1}^n M_i$ ($n\geq 1$) 
are examined under dimensional reduction to 4D-multiscalar 
fields \cite{zhuk}. In \cite{mign} a classification of MD 
inflationary models investigated.  
\par 
In this note the spherically symmetric solution in MD gravity 
is sought, which can be a domain, where 
ED is remain in ON-state even though 
compactification have taken place in our MD Universe. 
These would be regions with very strong 
gravitational fields. 
\par
In this work we examine such a MD gravity on a principal bundle. 
The importance of such theory follows from the following theorem 
\cite{wit}-\cite{sal}:
\par
Let $G$  group  be  the fibre of a principal  bundle.  Then  
there  is  the  one-to-one 
correspondence between $G$-invariant metrics on the  total  space
${\cal X}$ 
and the triplets $(g_{\mu \nu }, A^{a}_{\mu }, h\gamma _{ab})$, 
where $g_{\mu \nu }$ is Einstein's pseudo  - 
Riemannian metric, $A^{a}_{\mu }$ is a gauge field  of  the $G$  group, 
and $h\gamma _{ab}$ a symmetric metric on the fibre. In this case 
we can write down the MD Ricci scalar $R^{(MD)}$ in the following 
way:
\begin{equation}
 R^{(MD)} = R^{(4)} + R^{(G)}  -
 \frac{1}{4}F^a_{\mu\nu}F^{a\mu\nu}
 - d \partial _\mu (g^{\mu\nu}h^{-1}\partial _\nu h)-
  \frac{d(d+1)}{4h^2}\partial _\mu h \partial ^\mu h ,
\label{1-1}
\end{equation}
where $R^{(4)}$ is Ricci scalar of Einstein's 4D spacetime; 
$R^{(G)}$ is Ricci scalar of the gauge group $G$, 
$F^a_{\mu\nu} = \partial _\mu A^\nu - \partial _\nu A^a_\mu -
f_{abc}A^b_\mu A^c_\nu $ is the gauge field strength, 
$d$ is the dimension of 
the gauge group, $\nabla _\mu$ is the covariant derivative 
on 4D spacetime and $f_{abc}$ are the structural constants of given 
gauge group, $h$ is linear size of fibre and tensor Ricci 
$R_{\mu\nu} = R^\alpha_{\mu\nu\alpha}$. 
\par
This theorem tell us that the nondiagonal components of MD metric 
can be continued to 4D region as physical gauge fields 
(electromagnetic, U(2) or SU(3)). Of course we must have the 
compactification mechanism on the boundary between domains 
with compactified and noncompactified ED. 
We don't discuss this phenomenon but assume its existence.

\section{The gravity equation}

We note that the metric on the fibre has the following form: 
\begin{equation}
ds^{2}_{\rm fiber} = h(x^{\mu }) \sigma ^{a}\sigma _{a},
\label{2-1}
\end{equation}
where conformal factor $h(x^{\mu })$ depends only on spacetime 
coordinates $x^{\mu }$, here $\mu =0,1,2,3$ are the spacetime 
indexes, $\sigma _{a}=\gamma _{ab}\sigma ^{b}; \gamma _{ab}$  
is the euclidean  
metric and $a=4,5,\ldots \dim G$ index on fibre (internal space). 
This follows 
from the fact that the fibre is a symmetrical space (gauge group). 
$\sigma ^{a}$  are  one-forms which satisfy Maurer - Cartan 
structure equations: 
\begin{equation}
d\sigma ^{a} = f^{a}_{bc}\sigma ^{b}\wedge\sigma ^{c},
\label{2-2}
\end{equation}
where $f^{a}_{bc}$ is a structural constant of gauge group. 
Thus, MD metric on  the  total  space
can be written in the following view: 
\begin{equation}
ds^{2} = ds^{2}_{fibre} + 2 G_{A\mu } dx^{A} dx^{\mu },
\label{2-3}
\end{equation}
where $A=0,1,\ldots ,\dim G$  is  multidimensional  index 
on the total space. 
\par
Hence we have only following independent degrees of freedom: 
conformal factor $h(x^\mu)$ and MD metric $G_{A\mu}$. 
Varying with respect to these variables leads to the following 
gravity equations: 
\begin{eqnarray}
R^{(MD)}_{A\mu } -{1\over 2}G_{A\mu }R^{(MD)} & = & 0,
\label{2-2-1}\\
R^{(MD)a}_{\qquad a} & = & 0.
\label{2-2-2}
\end{eqnarray}
These equations are vacuum Einstein's MD equations for gravity on the 
principal bundle. 
\par
We can also write down these equations in 4D form using the Lagrangian
(\ref{1-1}):
\begin{equation}
  R^{(4)}_{\mu\nu}  - \frac{1}{2}g_{\mu\nu} R^{(4)} = T_{\mu\nu},
\label{2-5}
\end{equation}
where $T_{\mu\nu}$ is energy-impulse tensor for gauge field $A^a_\mu$ 
and scalar field $h(x^\mu)$.

\section{SU(3) spherically symmetric solution}

Ansatz for SU(3) potential we take the same as for SU(3) 
black hole in 4D gravity \cite{gal}:
\begin{eqnarray}
  \label{3-1-1}
  A_0 = T_a \frac{x^a}{r}\varphi(r) + \lambda _8 w(r),\\
  \label{3-1-2}
  A_i = T_a \epsilon _{aij}\frac{x^j}{r^2}(1 - f(r))
\end{eqnarray}
for embeddings of the SU(2) group into SU(3) with the following 
choices of generators $T_a$ in terms of Gell - Mann matrices:
\begin{equation}
  \label{3-2} 
  T_a = \frac{1}{2}(\lambda _1, \lambda _2, \lambda _3)
\end{equation}
this is for isospin $1/2$. And for isospin 1:
\begin{eqnarray}
  \label{3-3-1}
  A_0 = T_a \frac{x^a}{r}\varphi(r) + 
        \left (\frac{x^\alpha x^\beta}{r^2} -
        \frac{1}{3}\delta_{\alpha\beta}\right )
        w(r),\\
  \label{3-3-2}
  A_i = T_a \epsilon _{aij}\frac{x^j}{r^2}(1 - f(r)) + 
\left (\epsilon_{is\alpha}x_\beta + \epsilon_{is\beta }x_\alpha 
\right )\frac{x^s}{r^3}v(r),
\end{eqnarray}
here $\alpha , \beta =1,2,3$ are indexes of matric. 
Generators $T_a$ for isospin 1 are following:
\begin{equation}
  \label{3-4}
  T_a = (\lambda _7, -\lambda _5, \lambda _2).
\end{equation}
4D metric we search in following wormhole - like form:
\begin{equation}
  \label{3-5}
  ds^{2} = e^{2\nu (r)}dt^{2} - dr^{2}-
a^2(r)(d\theta ^{2} + \sin ^{2}\theta  d\phi ^2).
\end{equation}
This 4D metric and gauge field $A^a_\mu$ correspond to following 
MD metric:
\begin{eqnarray}
ds^{2} = e^{2\nu (r)}dt^{2}  - r^{2}_{0}e^{2\psi (r)}\sum^{8}_{a=1}
\left (\sigma ^{a} - A^{a}_{\mu }(r)dx^{\mu }\right )^{2} -
\nonumber \\
dr^{2} - a^{2}(r)\left (d\theta ^{2} + \sin ^{2}\theta d\phi ^2\right ).
\label{3-6}
\end{eqnarray}
For obtaining the field equations we write down the Euler 
equations for Lagrangian $\sqrt{-G}R^{(MD)}$ after substitution 
gauge field (\ref{3-1-1}) - (\ref{3-1-2}) and 4D metric (\ref{3-5}).
As we consider the vacuum Einstein's equations, we must 
write down $R^{(MD)}=0$ equation in addition. For simplicity 
we examine the case $\varphi (r)=0, f(r)=1$. 
Finally after some transformation we have the following 
system of equations:
\begin{eqnarray}
  \label{3-7-1}
  \nu '' + \nu ' \left (\frac{a'}{a} + \nu ' + 8 \psi ' \right ) - 
  \frac{1}{2} r_0^2 \exp{(2\psi - 2\nu)}{w'}^2 & = & 0,\\
  \label{3-7-2}
  \frac{a''}{a} - 2 + \frac{a'}{a} \left (\nu ' + 
  8 \psi ' \right ) & = & 0,\\
  \label{3-7-3}
  8\psi '' + 8\psi '\left (\frac{a'}{a} + \nu ' + 8 \psi ' \right )
  + \frac{1}{2}r_0^2 \exp{(2\psi - 2\nu)}{w'}^2 - 
  48\frac{\exp{(-2\psi)}}{r_0^2} & = & 0,\\
  \nonumber
  -2\frac{a'}{a}\left (\nu ' + 8\psi '\right ) - 16\psi '\nu ' - 
  56{\psi '}^2 + \frac{2}{a} - \frac{{a'}^2}{2a^2} & - & \\
  \label{3-7-4} 
  \frac{1}{2} r_0^2 \exp{(2\psi - 2\nu)}{w'}^2 + 
  24\frac{\exp{(-2\psi)}}{r_0^2} & = & 0, \\
  \label{3-7-5}
  \Biggl(\exp{(10\psi - \nu)} a w'\Biggr)' & = & 0, 
\end{eqnarray}
where $(\prime)$ means the derivative with respect to $r$. 
The last equation (\ref{3-7-5}) is ``Yang - Mills'' equation 
(nondiagonal Einstein's equation) which has the following 
solution:
\begin{equation}
  \label{3-8}
  w' = \frac{q}{ar_0} \exp{(\nu - 10\psi)},
\end{equation}
where $q$ is an integration constant (``color charge''). 
We consider the simplest case when ``color charge'' $q$ and/or 
size of extra dimension $r_0$ is very big:
\begin{equation}
  \label{3-9}
  \exp{(8\psi)} \ll \frac{qr_0}{a}.
\end{equation}
Then we have the following approximate equations:
\begin{eqnarray}
  \label{3-10-1}
  \nu '' + \nu ' \left (\frac{a'}{a} + \nu ' + 8 \psi ' \right ) - 
  \frac{q^2}{2r_0^2} \exp{(-18\psi)} & = & 0,\\
  \label{3-10-2}
  \frac{a''}{a} - 2 + \frac{a'}{a} \left (\nu ' + 
  8 \psi ' \right ) & = & 0,\\
  \label{3-10-3}
  8\psi '' + 8\psi '\left (\frac{a'}{a} + \nu ' + 8 \psi ' \right )
  + \frac{q^2}{2r_0^2} \exp{(-18\psi)} & = & 0,\\
  \label{3-10-4}
  -2\frac{a'}{a}\left (\nu ' + 8\psi '\right ) - 16\psi '\nu ' - 
  56{\psi '}^2 + \frac{2}{a} - \frac{{a'}^2}{2a^2} - 
  \frac{q^2}{2r_0^2} \exp{(-18\psi)} & = & 0,
\end{eqnarray}
These equations have the following solution
\begin{eqnarray}
  \label{3-11-1}
  \nu & = & -8\psi,
  \label{3-11-2}\\
  \exp{(9\psi)} & = & \frac{q}{2a_0}\cos {\left (\frac{3}{2}
  \arctan{\left (\frac{r}{a_0}\right )}\right )},\\
  \label{3-11-3}
  w & = & \frac{8}{3}\frac{a_0}{qr_0}\tan{\left (\frac{3}{2}
  \arctan{\left (\frac{r}{a_0}\right )}\right )},
\end{eqnarray}
where $a_0 = a(0)$. In this case the condition (\ref{3-9}) leads to the
following:
\begin{equation}
  \label{3-12}
  \frac{qr_0^9}{a_0^{10}} \gg 1.
\end{equation}
Let us define $r_H$ by which we will have the null surfaces 
$ds^2=0$:
\begin{equation}
\label{3-13}
g_{tt}(r_H) = \exp{(2\nu(r_H))} - r_0^2\exp{(2\psi(r_H))}\sum ^8_{a=1}
\left(A_t^2(r_H)\right) = 0.
\end{equation}
It is easy to see that this is may be satisfied 
by $r_H = \pm a_0$. By $r=0$ we have 
the throat of wormhole, hence we can say that this solution 
is a wormhole - like object located between two null surfaces 
$r_H = \pm a_0$.
\par 
Note that earlier the similar solutions was obtained 
for U(1) (5D gravity) \cite{dzh1} and for SU(2) 
(7D gravity) \cite{dzh2} cases.

\section{Discussion}
Thus, we can say that all MD gravity on the principal 
bundles with physical significant structural (gauge) 
groups U(1), SU(2) and SU(3) have the spherically symmetric 
wormhole - like solutions located between two null surfaces. 
In these theories gravity acts 
on whole total space of principal bundle (maybe this is 
a situation near gravitational singularity). If we suppose 
that the compactification mechanism of ED exists in nature then 
we can sew these wormhole - like objects with corresponding black holes: 
wormhole-like solutions for 5D, 7D and 12D gravities with 
Reissner - Nordstr\"om's black hole, SU(2) and SU(3) sphalerons 
respectively. The compactification mechanism based on 
algorithmical viewpoint and relevant for such joining 
in \cite{dzh3} is considered. Such composite wormholes will 
connect two asymptotically flat spaces. For U(1) case this is a 
model of electrical charge \cite{dzh4} without charge proposed by 
J.Wheeler.

\section{Acknowledgments}

I am very grateful prof.H.Kleinert for invitation 
to Freie Universit\"at Berlin and DAAD for stipendium.

\end{document}